\documentclass[%
 reprint,
 amsmath,amssymb,
 aps,
 prl,
]{revtex4-1}

\usepackage[utf8]{inputenc}
\usepackage{ae}
\usepackage{aecompl}

\usepackage{graphicx}
\usepackage{dcolumn}
\usepackage{bm}
 
\usepackage{siunitx}

\begin{document}

\title{Spin-isomer conversion of water at room temperature, and quantum-rotor-induced nuclear polarization, in the water-endofullerene H$_2$O@C$_{60}$} 

\author{Benno Meier}
\email{b.meier@soton.ac.uk}
\author{Karel Kou\v{r}il}
\thanks{B Meier and K Kou\v{r}il contributed equally.}
\author{Christian~Bengs}
\author{Hana Kou\v{r}ilov\'{a}}
\author{Timothy~J.~Barker}
\author{Stuart~J.~Elliott}
\author{Shamim Alom}
\author{Richard J. Whitby}
\author{Malcolm H. Levitt}
\email{mhl@soton.ac.uk}
\affiliation{School of Chemistry, University of Southampton, Southampton, SO17 1BJ, United Kingdom}

\newcommand{\cSixSeventeen}{H$_2{}^{17}$O@C$_{60}$}
\newcommand{\cSixSixteen}{H$_2{}^{16}$O@C$_{60}$}
\newcommand{\cSix}{H$_2$O@C$_{60}$}
 
\date{\today}

\begin{abstract} 
  Water exists in two forms, \textit{para} and \textit{ortho}, that have nuclear spin states with different symmetries. Here we report the conversion of fullerene-encapsulated \textit{para}-water to \textit{ortho}-water. The enrichment of \textit{para}-water at low temperatures is monitored via changes in the electrical polarizability of the material. Upon rapid dissolution of the material in toluene the excess \textit{para}-water converts to \textit{ortho}-water. In \cSixSixteen{} the conversion leads to a slow increase in the NMR signal. In \cSixSeventeen{} the conversion  gives rise to weak signal enhancements attributed to quantum-rotor-induced nuclear spin polarization. The time constants for the spin-isomer conversion of fullerene-encapsulated water in ambient temperature solution are estimated as 30$\pm\SI{4}{s}$ for the $^{16}$O-isotopologue of water, and 16$\pm\SI{3}{s}$ for the $^{17}$O isotopologue.
\end{abstract}

\pacs{Valid PACS appear here}
\maketitle




The wavefunction of a water molecule is antisymmetric under exchange of its two protons, and may be written as a product of a rotational state and a nuclear spin state. Two different spin isomers of water exist: In \textit{para}-water the rotational state is symmetric, with an antisymmectric nuclear spin state with total spin $I=0$. In \textit{ortho}-water the rotational state is antisymmetric, and the nuclear spin state is one of the three symmetric triplet states, with total spin $I = 1$.

The ratio of water spin isomers has been used  to estimate the formation temperature of water in astrophysics  \cite{encrenaz-2008-water-solar-system,hogerheijde-2011-detec-water}. Differences in the magnetic and electric properties have been used to separate the spin isomers in molecular beam experiments \cite{kravchuk-2011-magnet-focus, horke-2014-separ}. The conversion of water between its spin isomers has been followed at low temperatures by infrared spectroscopy \cite{redington-1963-molec-rotat, turgeon-2012-prepar-isolat} in inert gas matrices. A remarkable lifetime of 26 minutes has been claimed for bulk \textit{para}-water under ambient conditions  \cite{tikhonov-2002-separ-water}, although this claim has been disputed \cite{cacciani-2012-nuclear-spin, veber-2006-possib-enric}.

The molecular endofullerene \cSixSixteen{} has been produced by multi-step chemical synthesis \cite{kurotobi-2011-singl-molec, krachmalnicoff-2014-optim-scalab}, and provides an excellent system for the study of the water spin isomers. Each C$_{60}$ fullerene cage encapsulates a single water molecule which retains free molecular rotation at low temperatures  \cite{beduz-2012-quant-rotat, goh-2014-symmet-break}. The cage prevents the exchange of protons between water molecules, which provides a mechanism for rapid \textit{ortho}-\textit{para} equilibration in bulk water \cite{mammoli-2015-chall-prepar}. Spin-isomer conversion in water endofullerenes has been studied by infrared spectroscopy, neutron scattering, NMR, and via changes in dielectric constant \cite{beduz-2012-quant-rotat,goh-2014-symmet-break,mamone-2014-nuclear-spin,meier-2015-elect-detec}. The conversion of \textit{ortho}- to \textit{para}-water follows second order kinetics \cite{mamone-2014-nuclear-spin}, indicating that it is facilitated by water-water interactions.

\begin{figure}[b]
  \includegraphics{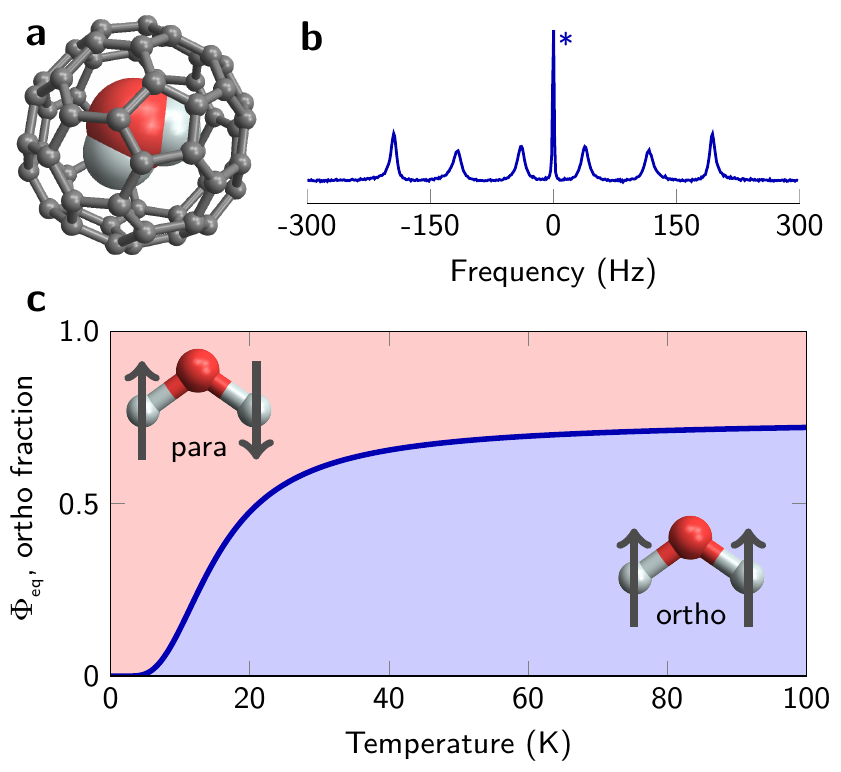}\vspace{-3mm}
  \caption{(a) Molecular structure of \cSix. (b) $^{1}$H NMR spectrum of \cSix, with a $^{17}$O labelling level of 85$\,$\%, dissolved in deuterated ortho-dichlorobenzene. The proton resonance is split into a sextet due to the $J$-coupling to the spin-$5/2$ oxygen. The central line is due to the 15\% of water bearing the $^{16}$O isotope. (c) Equilibrium \textit{ortho}-fraction as a function of temperature. At high temperatures, 85\% of the molecules are in the ortho state. At a temperature of 5~Kelvin and in thermal equilibrium, the \textit{ortho}-state is almost completely depleted.\label{fig:introduction}}
\end{figure}

Here we report on the conversion between \textit{para} and \textit{ortho} water spin isomers in the endofullerene \cSix, in room-temperature solution. 
Alongside the spin-isomer conversion, we also observe weak nuclear spin polarization effects for water molecules bearing a $^{17}$O nucleus. These quantum-rotor-induced polarization (QRIP) effects are related to analogous effects observed for the case of unhindered $^{13}$C-bearing methyl (CH$_3$) rotors \cite{icker-2012-unexp-multip,icker-2013-exper-bound,meier-2013-long-lived}. 


The $^{1}$H NMR spectrum of H$_2$O@C$_{60}$, with a $^{17}$O labelleling level of 85$\,$\% and dissolved in deuterated ortho-dichlorobenzene (ODCB), is shown in Fig.~\ref{fig:introduction}. The $^{1}$H NMR spectrum exhibits 6 peaks due to the $J$-coupling of the two equivalent protons to the spin-5/2 $^{17}$O nucleus. The linewidth differences of the $^{1}$H NMR peaks are due to $^{17}$O quadrupolar relaxation \cite{elliott-2017-nmr-lines}. The sharp central peak is due to the 15$\,$\% of molecules bearing the $^{16}$O isotope. At 25$^\circ$ the $^{1}$H $T_1$ values were reported previously to be $700\pm\SI{50}{ms}$ for \cSixSixteen{}, and $750\pm\SI{50}{ms}$ for \cSixSeventeen. The $^{17}$O $T_1$ is $81\pm\SI{7}{ms}$ \cite{elliott-2017-nmr-lines}.  

The \textit{ortho}-H$_2$O ground state in \cSix{} is approximately \SI{2.7}{meV} higher in energy than the \textit{para}-H$_2$O ground state, as determined by neutron scattering \cite{beduz-2012-quant-rotat}. Hence temperatures below $\sim$\SI{20}{K} strongly favour the \textit{para}-H$_2$O state in thermal equilibrium. Fig.~1c shows the equilibrium proportions of \textit{ortho} and \textit{para}-H$_2$O as a function of temperature.

In the current work, we enhanced the fraction of \textit{para}-H$_2$O by thermal equilibration at low temperature, and raised the temperature rapidly by dissolving the material in a warm solvent. The $^{1}$H NMR spectrum was observed while \textit{para}-water converts back to \textit{ortho}-water at ambient temperature.

One experimental difficulty is that solid H$_2$O@C$_{60}$, just like C$_{60}$, dissolves only slowly, due to strong intermolecular interactions. 
Since slow dissolution is incompatible with observation of the rapid \textit{para}-to-\textit{ortho} conversion process, we first dissolved H$_2$O@C$_{60}$ in deuterated ODCB, at a concentration of \SI{17}{mM} \cite{marcus-2001-solub-c60ful}. Dissolution proceeds rapidly when the frozen solid solution is exposed to an excess of warm solvent.

The kinetics of \textit{ortho}-\textit{para} conversion in pure H$_2$O@C$_{60}$ solid indicate that interactions between water molecules are important for the spin-isomer conversion process \cite{mamone-2014-nuclear-spin}. The use of a frozen solution generates uncertainty as to whether the water molecules in the frozen endofullerene solution do become enriched in the \textit{para} spin isomer.  
To resolve this issue we studied the \textit{ortho}-\textit{para} conversion of \cSixSixteen{} in frozen  ODCB solution %
%
by measuring the dielectric constant of the material \cite{meier-2015-elect-detec}. The electric polarizability of \textit{ortho}-water is larger than that of \textit{para}-water, in their respective rotational ground-states. As water converts from \textit{ortho}- to \textit{para}-water, the dielectric constant decreases. In solid \cSixSixteen{}, this process takes many hours to complete and is well described by second order kinetics \cite{mamone-2014-nuclear-spin}.

Solutions of 25 mg of C$_{60}$, \cSixSixteen{} and \cSixSeventeen{} in 2 mL ODCB were prepared. \SI{600}{\micro\liter} of each solution were injected into one of three capacitors of a home-built capacitance probe. Details of the apparatus are given in the supporting information (SI).  To allow time for the \textit{ortho}-to-\textit{para} conversion, the temperature was stabilized at 25 Kelvin for \SI{10}{h} using a commercial flow cryostat. Subsequently the temperature was reduced within \SI{6}{min} to \SI{5.05}{K} where it was stabilized with an accuracy of $\pm\SI{30}{mK}$. Capacitance data were recorded approximately every \SI{30}{s} for all three capacitors during the \SI{10}{h} prior to the temperature drop, and for \SI{28}{h} after the temperature drop.

\begin{figure}[t]
  \includegraphics{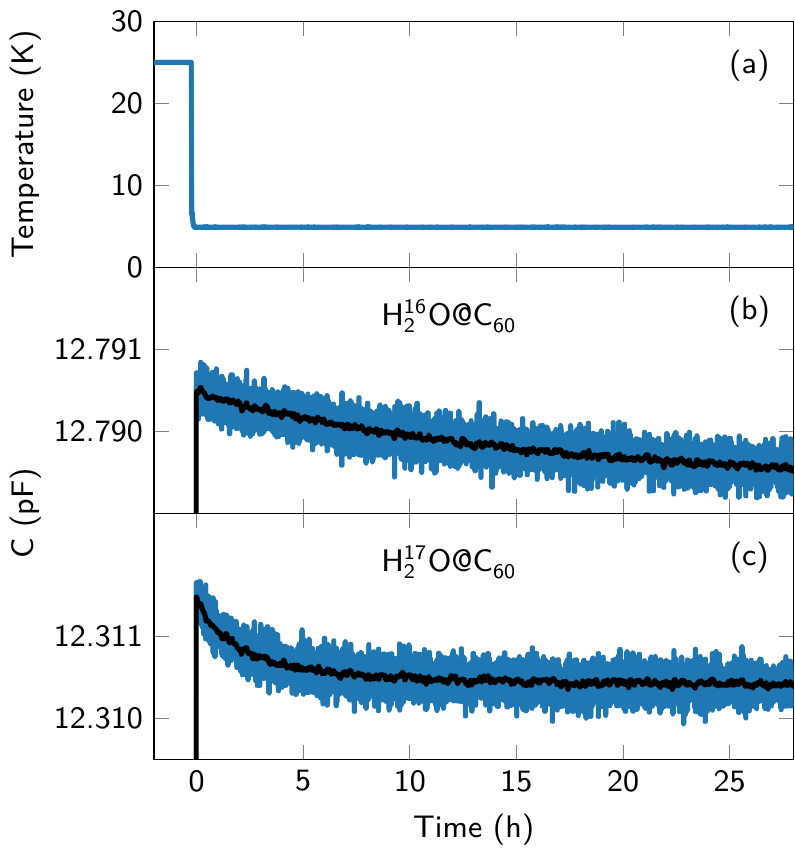}\vspace{-3mm}
  \caption{Dielectric detection of \textit{ortho}-\textit{para} conversion of fullerene encapsulated water dissolved in ODCB. (a) Temperature profile. The sample is kept at 25 Kelvin for 10 h, followed by a temperature drop to 5 Kelvin. The time origin coincides with the temperature drop. (b) Capacitance of the capacitor filled with H$_2^{16}$O@C$_{60}$. (c) Capacitance of the capacitor filled with \cSixSeventeen. All capacitance values were corrected for temperature fluctuations (see SI).   The initial increase in capacitance is due to increased polarizability of both water and the host.  A subsequent decay due to ortho-para conversion is apparent in both samples, but proceeds substantially faster in $^{17}$O-labelled water. The black curves in (b) and (c) were obtained by applying a local bandwidth reduction of width \SI{5}{min}.\label{fig:capacitance}}
 \end{figure}


 The capacitance data of the \cSixSixteen{} and \cSixSeventeen{} capacitors were corrected for small thermal fluctuations using data measured on the C$_{60}$ capacitor. The temperature profile and corrected data are shown in Fig.~\ref{fig:capacitance}.  The black lines in panels (b) and (c) are obtained using a locally weighted regression (LOESS) \cite{janert-2010-data-analy-with} with a tri-cube kernel $k(t) = (1 - |t/b|^3)^3$ of bandwidth $b =  \SI{5}{min}$. Details are given in the SI.

The rapid temperature drop leads to an increase in raw capacitance data of approximately \SI{10}{fF} for all three capacitors. During the subsequent \SI{28}{h}, the capacitances of the $^{16}$O and $^{17}$O capacitors drop slowly by 0.9 and 1.0 fF, respectively. This change is attributed to \textit{ortho}/\textit{para} conversion. As can be seen in Fig.~\ref{fig:capacitance}, the \textit{ortho}-\textit{para}-conversion proceeds faster in \cSixSeventeen{} than in \cSixSixteen{}. As detailed in the SI, the \cSixSeventeen{} data are best fit using a second order model. The more slowly decaying \cSixSixteen{} data, on the other hand, fit a second order model and an exponential decay equally well. Faster \textit{ortho}/\textit{para}-conversion of the $^{17}$O labelled water has also been observed on solid samples of pure \cSixSixteen{} and \cSixSeventeen{}. Corresponding data are shown in the SI.

Previous studies of nuclear spin conversion in \cSixSixteen{} by means of NMR and dielectric measurements have shown that the material does not fully equilibrate - only a fraction of the \textit{ortho}-water molecules converts to \textit{para}-water. This fraction may be estimated from the change in capacitance using the Clausius-Mossotti equation. For both the $^{16}$O and the $^{17}$O capacitor the dielectric constant is calculated by taking the ratio of the corrected capacitance and the measured capacitance of the empty capacitor. Thus determined values of the dielectric constant $\epsilon$ are used to calculate the bulk electrical polarizability $\mathcal{P}$, using the Clausius-Mossotti equation:
\begin{equation}
  \mathcal{P} = 3 \epsilon_0 \frac{\epsilon -1}{\epsilon + 2}. 
\end{equation}

The changes in $\mathcal{P}$ are attributed to changes of the electrical polarizability of the water molecules as they convert from \textit{ortho} to \textit{para}. The average change in polarizability of the water molecules, $\Delta \alpha$, may be calculated as $\Delta \alpha = \Delta \mathcal{P} / N$, where $N$ is the known volume density of the water molecules. For $^{16}$O-water and $^{17}$O-water we find changes in average polarizability volume $\Delta \alpha ' = \Delta  \alpha / (4 \pi \epsilon_0)$ of 0.9 and 1.2 \AA$^3$ over \SI{28}{h} during which the samples are kept at 5 Kelvin.

Previous experimental and theoretical estimates have found that the polarizabilities of \textit{ortho} and \textit{para}-water in their respective rotational ground states differ by 12 $\pm$ 3 \AA$^3$ \cite{meier-2015-elect-detec, chang-2014-cmist}.  The capacitance data therefore indicate that only 5 to 13\% of the water molecules present in the sample convert from \textit{ortho} to \textit{para} during the time the sample is kept at 5 K. Since the extent to which the sample has reached thermal equilibrium at 25 Kelvin is not known, this figure leads to an estimate for the final \textit{para}-fraction of 0.4$\,\pm\,0.1$. 

\begin{figure}
  \includegraphics{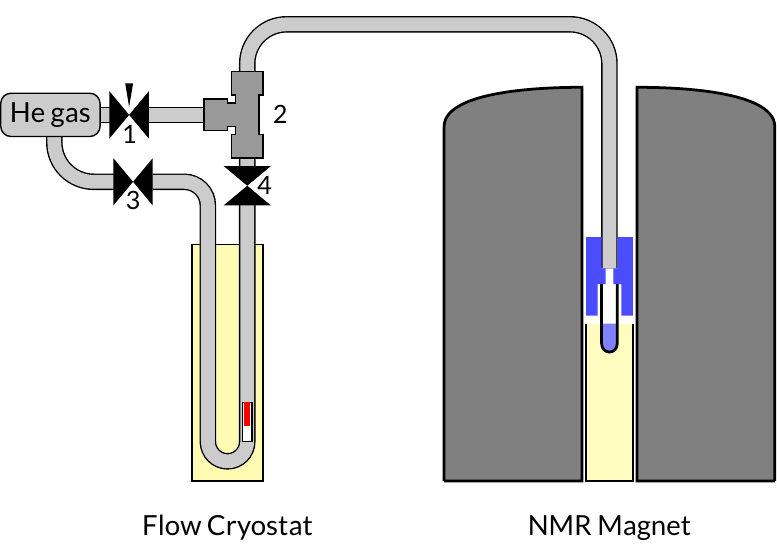}\vspace{-3mm}
  \caption{\label{fig:setupBullet} Apparatus for rapid dissolution. The sample (red) is pipetted into a teflon bullet which is then immersed in liquid nitrogen. For loading, the tube at the top of the union (2) is disconnected, the ball valve (4) is opened and the bullet is pushed to the bottom of the steel tube. A small flow of helium gas is applied via the needle valve (1) to prevent contamination of the system with air. The capsule is ejected by opening valves (3) and (4) and is propelled into the NMR magnet (right), where a 3D printed receiver (blue) retains the capsule. The sample travels further into the NMR tube where it dissolves in warm solvent.}
\end{figure}

In order to observe spin isomer conversion at ambient temperature, the sample is dissolved and the spectra are recorded using a liquid-state NMR apparatus. We rapidly dissolved the sample using a home-built penumatic shuttle, shown in Fig.~\ref{fig:setupBullet}. A sample volume of approximately \SI{50}{\micro\liter} is pipetted into a small teflon ``bullet''. The bullet is immersed in liquid nitrogen and then loaded into a U-shaped steel-tube that is fitted into a commercial flow cryostat, mounted, for an unrelated purpose, in a 6.7 Tesla magnet. The field has no significant effect on the \textit{ortho}/\textit{para} conversion, but leads to a magnetization of the \textit{ortho}-water. After keeping the sample at low temperature for the desired time, valves are opened on both terminals of the steel tube, and the capsule is shot out using pressurized helium gas at 8 bar. The transfer of the sample to the secondary magnet takes approximately \SI{100}{ms}. Inside the high-field magnet, a 3D printed receiver structure vents the helium gas, retains the sample capsule, but allows for the sample to slide out of the capsule into a \SI{5}{mm} NMR tube pre-loaded with \SI{0.5}{mL} of solvent. This method allows dissolution with a much smaller volume of solvent than previous procedures \cite{ardenkjaer-larsen-2003-increas-signal}.

\begin{figure}[b]
  \includegraphics{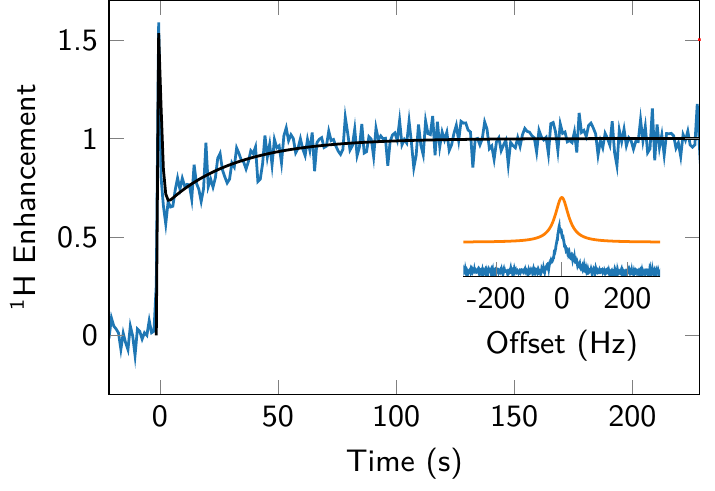}\vspace{-3mm}
  \caption{Intensity of the $^1$H NMR signal of \cSixSixteen{} in solution as a function of time (blue curve). Acquisition starts prior to arrival of the sample. The time origin is chosen to coincide with the first NMR signal after dissolution of the sample. The initial sharp peak is due to initial \textit{ortho}-water magnetization (the cryostat is located in a 6.7 Tesla magnet), and decays rapidly with the proton $T_1$. Following the rapid decay the signal intensity increases with time as \textit{para}-water converts to \textit{ortho}-water. The black curve shows the result of a spin dynamical simulation with a time constant $T_{PO}$ for the equilibration of \textit{ortho}- and \textit{para}-water of \SI{30}{s} (see text). The inset shows a spectrum (lower, blue curve) obtained from averaging 20 transients recorded 8 min after dissolution of the sample, and a Lorentzian mask (upper, orange curve) that was applied prior to integration. \label{fig:nmr16}}
\end{figure}
 
\begin{figure*}
  \includegraphics{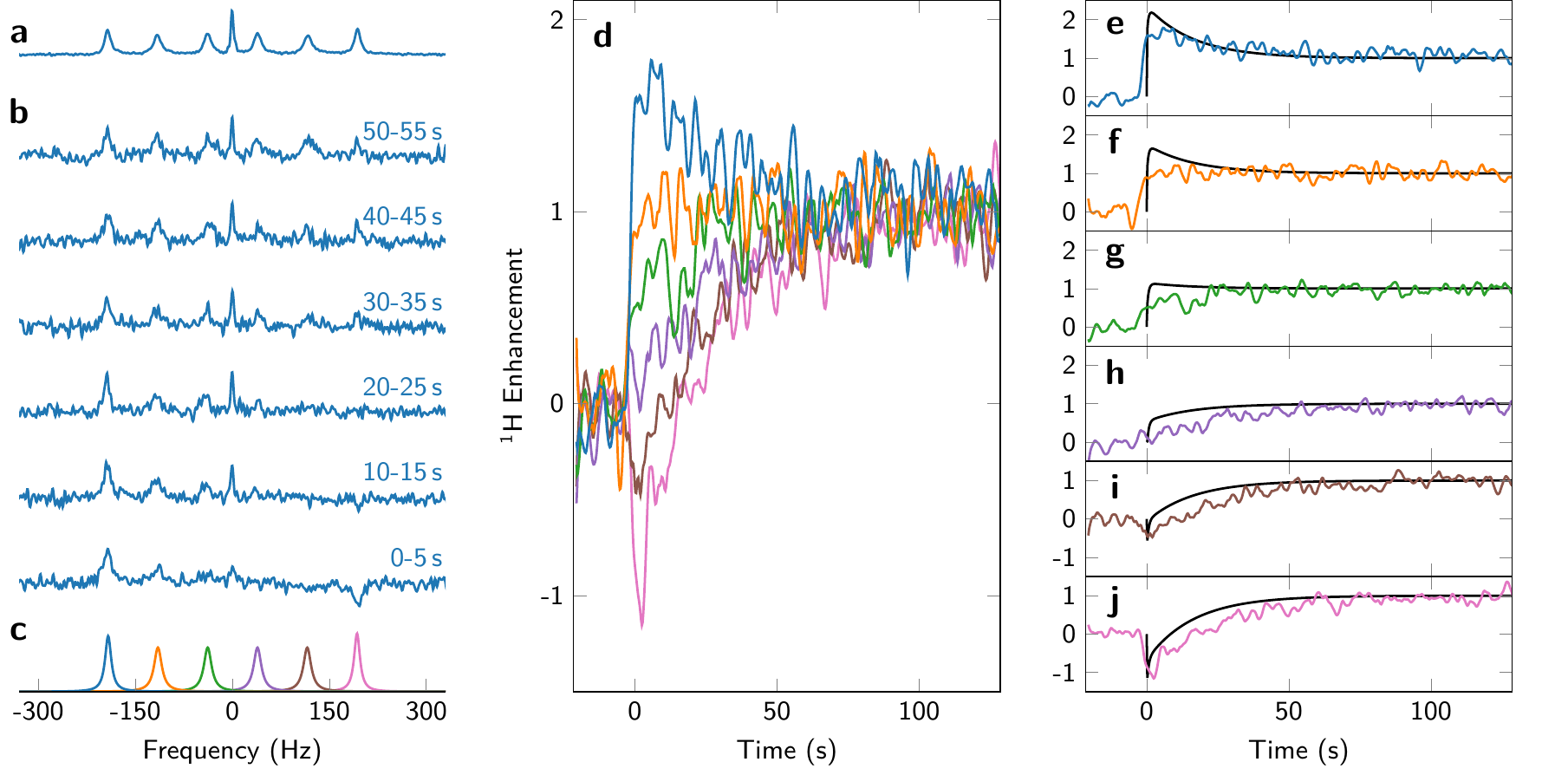}\vspace{-3mm}
  \caption{\label{fig:nmr17} QRIP Experiment on \cSixSeventeen{}. (a) Thermal equilibrium $^{1}$H spectrum, recorded at the end of the experiment (1024 transients averaged). (b) Averages of 20 transients during consecutive intervals after dissolution of the sample. A clear antiphase signal is obtained immediately after the temperature drop (0-5s). The antiphase signal decays, and after approximately one minute the thermal signal is restored. (c) The individual spectra are multiplied with Lorentzian masks that correspond to each of the six peaks, and the product is integrated. (d) Results of the integration after applying a tri-cube bandwidth reduction of width \SI{2}{s}. The integrals are normalized by the intensity of the respective thermal $^{17}$O peaks. (e-j) The same data as in (d), with spin dynamical simulations (black lines).}
\end{figure*}

Separate experiments were conducted to study the \textit{para}/\textit{ortho} conversion of \cSixSixteen{} and \cSixSeventeen{}. In the first experiment a sample of \SI{50}{\micro\liter} \cSixSixteen{} in ODCB was kept at a temperature of 3.9 Kelvin for approximately 15 hours. $^{1}$H NMR spectra  were recorded every \SI{1}{s} using a flip-angle of approximately \SI{30}{\degree}. Acquisition of NMR data was started approximately \SI{20}{s} before the sample was ejected from the cryostat and dissolved in the NMR tube. The obtained spectra were multiplied with a Lorentzian mask and integrated to get the NMR signal intensity, shown in Fig.~\ref{fig:nmr16}, where the time origin has been chosen to coincide with the first NMR signal after dissolution of the sample. The inset of Fig.~\ref{fig:nmr16} shows an approximate thermal equilibrium spectrum, obtained by averaging 20 transients recorded 8 min after sample dissolution (lower, blue curve), and the Lorentzian mask (upper, orange curve).

The sharp peak in NMR signal immediately after dissolution is attributed to the nuclear magnetization of \textit{ortho}-water which was pre-polarized by keeping the sample in a magnetic field at low temperatures. This peak decays rapidly with the proton $T_1$ of $\sim \SI{0.8}{s}$. Subsequently the signal intensity slowly increases as \textit{para}-water converts to \textit{ortho}-water. The black curve in Fig.~\ref{fig:nmr16} shows the result of a simulation that yields a time constant $T_{PO} = \SI{30}{s}$ for the equilibration of \textit{ortho}- and \textit{para}-water (see below).

Results of an analogous experiment, conducted on \cSixSeventeen{}, are shown in Fig.~\ref{fig:nmr17}. For this experiment, the sample was kept at a temperature of 4 Kelvin for approximately 15 h. $^{1}$H NMR spectra were recorded every \SI{250}{ms} using a flip-angle of approximately \SI{30}{\degree}.

Figure~\ref{fig:nmr17}(b) shows the evolution of the NMR spectrum during the first minute after dissolution of the sample. Spectra were averaged over a total acquisition time of 5 seconds to improve the signal-to-noise ratio. Spectra averaged during the first 5 seconds clearly show an anti-phase pattern that is characteristic for QRIP \cite{meier-2013-long-lived,dumez-2015-theor-long,roy-2015-enhan-quant}. The time-evolution of the individual peaks is obtained by multiplying the spectra with Lorentzian masks as shown in panel (c), followed by integration. The corresponding trajectories for the six \cSixSeventeen{} peaks, normalized by their respective thermal intensities and with a tri-cube bandwidth reduction of width \SI{2}{s}, are shown in panel (d). Panels (e-j) show separate plots of the six peak intensities, together with trajectories obtained from spin dynamical simulations.  The outermost peaks exhibit small enhancements of up to 2. A thermal equilibrium spectrum, recorded after full equilibration of the sample (average of 1024 transients) is shown in Fig.~\ref{fig:nmr17}(a).

Simulations of the proton peak trajectories during the spin-isomer conversion process of both \cSixSixteen{} and \cSixSeventeen{} were performed by using {\em SpinDynamica} software \cite{bengs-2017-spind}. Since \textit{para}-water has a singlet nuclear spin state, and \textit{ortho}-water has a triplet nuclear spin state, the simulation methodology is analogous to that used for studying the dynamics of long-lived nuclear singlet states \cite{carravetta-2004-beyon-thet1, pileio-2016-relax-theor,levitt-2012-singl-nuclear,stevanato-2015-nuclear-singl}. 
The simulations for H$_2{}^{16}$O included the following relaxation mechanisms: $^{1}$H-$^{1}$H dipole relaxation and spin-rotation. In the case of H$_2{}^{17}$O,  $^{1}$H-$^{17}$O dipole-dipole relaxation, $^{17}$O quadrupolar relaxation, and their cross-correlations were also included, along with the experimentally observed  $^{1}$H-$^{17}$O $J$-coupling of 77.9 Hz.

The relevant interactions were derived from the known molecular geometry and the $^{17}$O quadrupole coupling constant of \SI{10.11}{MHz} \cite{olsen-2002-elect-field}. In order to match the experimentally observed relaxation time of $86$ ms for the $^{17}$O longitudinal magnetization, a rotational correlation time of $\tau_C =\SI{107}{fs}$ was used for the simulation of both water isomers \cite{elliott-2017-nmr-lines}.

An external random field contribution was included in order to model spin-rotation coupling. All relaxation processes were treated by using a relaxation superoperator in the isotropic fast-motion limit as detailed in Ref.~\cite{elliott-2017-nmr-lines}. 

The random field contribution to the relaxation superoperator may be characterized by the magnitude of the random fields and correlation coefficients $\kappa_{i,j}$ as shown in~\cite{dumez-2015-theor-long}. A random field of ${\gamma^{\textrm{rand}^2}_{\textrm{H}}} \tau^{\textrm{rand}}$ of $0.6$ $\textrm{s}^{-1}$ was used in order to match the experimentally observed longitudinal relaxation time of \SI{755}{ms} for the proton spins.

The time constant for the \textit{para}/\textit{ortho} equilibration is related to the rates for \textit{para}-{to}-\textit{ortho} conversion, $k_{p\rightarrow o}$ and the reverse process, $k_{o \rightarrow p}$, as follows:
\begin{equation}
  T^{-1}_{PO} = 4 k_{p \rightarrow o} = \frac{4}{3} k_{o \rightarrow p}.
\end{equation}
The relaxation superoperator yields the following expressions for $k_{p \rightarrow o}$ for the two water isotopologues:
\begin{align}
 k_{p \rightarrow o} (\rm{H}_2{}^{16}\rm{O}) & = {\gamma^{\textrm{rand}^2}_{\textrm{H}}} \tau^{\textrm{rand}} (1 - \kappa)\\
 k_{p \rightarrow o} (\mbox{H}_2{}^{17}\mbox{O}) & = k_{p \rightarrow o} (\mbox{H}_2{}^{16}\mbox{O}) + k_{\mbox{DD}}^{\rm{OH}},
\end{align}
In the case of H$_2{}^{17}$O, $k_{\textrm{DD}}^{\rm{OH}}$ accounts for the additional contribution of the $^{1}$H-$^{17}$O dipole-dipole interactions. For $^{16}$O, good agreement with experimental data is found for $T_{PO} (\rm{H}_2{}^{16}\rm{O}) = 30\pm\SI{4}{s}$, corresponding to $\kappa = 0.986 \pm 0.002$. In the case of H$_2{}^{17}$O, the known molecular geometry and rotational correlation time $\tau_C$ leads to the following estimate of the heteronuclear dipole-dipole contribution:  $k_{\rm{DD}}^{\rm{OH}} = \SI{0.0093}{s}^{-1}$. Good agreement is found for $\kappa \in [0.99, 1]$, corresponding to $T_{PO} (\rm{H}_2{}^{17}\rm{O}) = 16\pm\SI{3}{s}$. This analysis shows that the experimental data and theory may only be reconciled by assuming a very high degree of correlation for the random fields used to model the spin-rotation interaction. 

In NMR, the evolution of the spin density operator is typically described by the following master equation \cite{ernst-1987-princ-nuclear}:
\begin{equation}
  \dot{\rho} (t) = -{i\mkern1mu}\hat{\textrm{H}}_{\mbox{\tiny coh}}\rho(t) +  \hat{\Gamma}(\rho(t) - \rho_{\mbox{\tiny eq}})
\end{equation}
where $\hat{\textrm{H}}_{\mbox{\tiny coh}}$ is the commutation superoperator of 
the coherent Hamiltonian and $\rho{\mbox{\tiny eq}}$ the thermal equilibrium density operator. This equation is a good approximation for spin systems which are always close to thermal equilibrium, but leads to incorrect results for spin systems that are far from equilibrium, which is the case here. The simulations shown in Figs.~\ref{fig:nmr16} and~\ref{fig:nmr17}  made use of the homogeneous master equation \cite{jeener-1982-super-opera,levitt-1992-stead-state}:
\begin{equation}
  \dot{\rho} (t) = -{i\mkern1mu}\hat{\textrm{H}}_{\mbox{\tiny coh}}\rho(t) + \hat{\Gamma}_{\theta}(\rho(t))
\end{equation}
The thermally corrected relaxation superoperator $\hat{\Gamma}_{\theta}$ is given by:
\begin{equation}
\hat{\Gamma}_{\theta} = \hat{\Gamma}\exp\left ( \hat{\textrm{H}}^{\textrm{L}}_{\mbox{\tiny lab}}\hat{P}_{D} / (k_\textrm{\tiny B} T)\right )
\end{equation}
with $\hat{\textrm{H}}^{\textrm{L}}_{\mbox{\tiny lab}}$ being the left translation superoperator \cite{ernst-1987-princ-nuclear} of the laboratory-frame coherent Hamiltonian $\textrm{H}_{\mbox{\tiny lab}}$ and $\hat{P}_{D}$ is the diagonal part projector with respect to the Hamiltonian $\textrm{H}_{\mbox{\tiny lab}}$ \cite{levante-1995-homog-versus}. Unlike the forms proposed in refs.  \cite{jeener-1982-super-opera,levitt-1992-stead-state}, this superoperator respects detailed balance \cite{levante-1995-homog-versus} and provides physically reasonable results even for spin systems far from thermal equilibrium.

The density operator immediately after dissolution is determined by the initial fraction of \textit{para}- and \textit{ortho}-water. 
The initial fractions are not known a priori and were adjusted to give the best agreement between simulations and data. The simulation results are shown as black curves in Fig.~\ref{fig:nmr16} and Fig.~\ref{fig:nmr17}(e-j) for \cSixSixteen{} and \cSixSeventeen{} respectively.  The initial \textit{para}-fractions are estimated as 54$\pm 15\,$\% and 43$\pm 3\,$\%, respectively, where the larger uncertainty for \cSixSixteen{} is due to the smaller dynamic range of the H$_2{}^{16}$O data. 

In conclusion, NMR experiments were used to study the spin isomer conversion for the endohedral water molecules of \cSix{} in ambient-temperature solution. The time constants for the \textit{para}/\textit{ortho} equilibration are found to be 30$\pm\SI{4}{s}$ for the $^{16}$O isotopologue of water, and 16$\pm\SI{3}{s}$ for the $^{17}$O isotopologue. The faster \textit{para}-to-\textit{ortho} conversion of H$_2{}^{17}$O is attributed to $^{1}$H-$^{17}$O dipole-dipole interactions. Weak nuclear spin polarization effects are observed for the $^{17}$O isotopologue as the spin isomer conversion proceeds. Measurements of electrical capacitance at low temperature indicate that slow conversion between the water spin isomers also takes place in frozen solutions of \cSix{} in ODCB, with the conversion again proceeding more rapidly for the $^{17}$O isotopologue than for the $^{16}$O species.

We acknowledge discussions with Walter K\"{o}ckenberger and Sankeerth Hebbar.
This research was supported by EPSRC (UK), grant
codes EP/N002482, EP/M001962/1, EP/L505067/1, and  EP/P009980 and the Wolfson Foundation.


%

\end{document}